\documentclass[useAMS,usenatbib]{mn2e}

\usepackage{amsfonts,amsmath,amssymb,bm,epsfig,url}
\usepackage[dvips]{color}

\newcommand{\df}[2]{ \frac{\partial {#1}}{\partial {#2}} }

\allowdisplaybreaks

\title[Magnetic field effects in neutron stars]{Consistent neutron
  star models with magnetic field dependent equations of state}

\author[D. Chatterjee, T. Elghozi, J. Novak and M. Oertel]{
 Debarati Chatterjee,$^1$\thanks{E-mail: debarati.chatterjee@obspm.fr} 
 Thomas Elghozi,$^2$\thanks{E-mail: thomas.elghozi@kcl.ac.uk}
 J\'er\^ome Novak$^1$\thanks{E-mail: jerome.novak@obspm.fr}
 and Micaela Oertel$^1$\thanks{E-mail: micaela.oertel@obspm.fr}\\
 $^1$Laboratoire Univers et Th\'eories, Observatoire de
 Paris, CNRS, Universit\'e Paris Diderot, 5 place Jules Janssen,
 92190 Meudon, France\\
 $^2$Department of Physics, King's College London, University of
 London, Strand, London WC2R 2LS, United Kingdom
}

\date{\today}

\pagerange{\pageref{firstpage}--\pageref{lastpage}} \pubyear{2014}

\begin{document}
\label{firstpage}

\maketitle

\begin{abstract}
  We present a self-consistent model for the study of the structure of
  a neutron star in strong magnetic fields. Starting from a
  microscopic Lagrangian, this model includes the effect of the
  magnetic field on the equation of state, the interaction of the
  electromagnetic field with matter (magnetisation), and anisotropies
  in the energy-momentum tensor, as well as general relativistic
  aspects. We build numerical axisymmetric stationary models and show
  the applicability of the approach with one example quark matter
  equation of state (EoS) often employed in the recent literature for
  studies of strongly magnetised neutron stars. For this EoS, the
  effect of inclusion of magnetic field dependence or the
  magnetisation do not increase the maximum mass significantly in
  contrast to what has been claimed by previous studies.
\end{abstract}


\section{Introduction}\label{s:intro}

One of the densest objects in the universe that can be observed
directly, neutron stars are perfect testing grounds for theories of
extreme physics. As the densities in the interior reach about $10^{15}
{\rm g.cm}^{-3}$, corresponding to several times normal nuclear matter
saturation density, the theoretical models describing cold and dense
matter, calibrated around the nuclear saturation point ($\sim 10^{14}
{\rm g.cm}^{-3}$) for symmetric nuclear matter (same number of protons
and neutrons), must be extrapolated in density as well as
asymmetry. To test these models, one can calculate the structure of
neutron stars by using the energy momentum tensor and solving
equations for hydrostatic equilibrium, and then compare them with
astrophysical observations. One constraint that these models should
satisfy is to be able to explain the highest observed neutron star
mass, which is $\sim 2 M_{\odot}$ according to the recent
astrophysical reports, see \cite{Demorest,Antoniadis}.

Neutron stars are not only extremely dense objects, but they are known
to be associated with strong magnetic fields, too. From pulsar
spin-down rates, employing the simple magnetic dipole model, the
estimated surface magnetic field value is typically $\sim
10^{12}-10^{13}$~G. Further, a few X-ray dim isolated neutron stars
(XDINSs) and rotating radio transients (RRATs) have recently been
observed with even higher magnetic fields, see \cite{Popov}. The
highest magnetic fields in neutron stars have been reported in soft
gamma-ray repeaters and anomalous X-ray pulsars. The common properties
of these two classes of objects have led to the proposition of a
unified model of magnetars to explain the observed features. Various
observations of magnetars, including the direct observation of
cyclotron lines (\cite{Ibrahim, Mereghetti}), indicate a magnetic
field value of up to $\sim 10^{15}$ G on the surface.

As the maximum magnetic field in the interior of magnetars cannot be
directly measured, it is generally estimated using the virial theorem
- most estimates point towards a maximal theoretical value $\sim
10^{18}$ G. Presence of a strong magnetic field can affect neutron
stars in two ways. Firstly, inclusion of the magnetic field results in
a modification of the energy momentum tensor, breaking the spherical
symmetry of the star and resulting in an anisotropy in the
latter. Secondly, it affects the Equation of State (EoS) due to Landau
quantization of the constituent particles, as pointed out in
\cite{DebaPRL}. One would thus expect the EoS as well as observational
quantities, such as the maximum mass, to be affected by strong
magnetic fields.

Previous models of magnetars, which included magnetic field effects on
the EoS, computed the corresponding mass-radius relations using
isotropic Tolman-Oppenheimer-Volkoff (TOV) equations, see e.g.~
\cite{Rabhi2009,Ferrer2010,Paulucci,Strickland2012,Lopes,Dexheimer,
  Casali}. Recently \cite{Ritam} attempted to compute the structure of
neutron stars in strong magnetic fields by a simple Taylor expansion
of the energy-momentum tensor and the metric around the spherically
symmetric case. However, it must be noted that at strong magnetic
fields for which Landau quantization effects start to become
non-negligible, the deviations from spherical symmetry are
significant. There exist global models of neutron stars in which the
effect of magnetic fields on the structure of neutron stars have been
included in a fully general relativistic formalism,
see~e.g.~\cite{Bocquet, Cardall, Oron, Kunihito, Kiuchi, Yasutake,
  Frieben, Yoshida}. These simulations demonstrate that the deviation
from spherical symmetry under strong magnetic fields is
remarkable. However for these numerical studies the effect of magnetic
field on the neutron star EoS was not taken into account.

In this paper, we present a self-consistent model to study neutron
stars with strong magnetic fields, including the effect of the
magnetic field on the EoS, general relativistic aspects as well as the
anisotropy of the energy-momentum tensor caused by the breaking of
spherical symmetry by the electromagnetic field and magnetisation. We
solve Einstein's equations in an axisymmetric metric which is
determined from the axisymmetric energy momentum tensor. We
investigate the effect of a strong quantizing magnetic field for the
particular case of quark matter in MCFL (Magnetic
Colour-Flavor-Locked) phase.

The paper is organised as follows. In Sec.~\ref{s:magtmunu} we
construct from the Lagrangian for fermions in an (electro)-magnetic
field the microscopic energy-momentum tensor. We calculate the
thermodynamic average of the energy-momentum tensor and identify the
contributions due to the magnetic field. In Sec.~\ref{s:global}, we
then write down the modified Maxwell equations and derive the
hydrodynamic equations in presence of a magnetic field. In
Sec.~\ref{s:resu}, we report our results and discuss their
consequences. Finally in Sec.~\ref{s:conc}, we summarise our findings
and conclusions. Some details of the derivations are provided in the
Appendix. Unless otherwise stated, we work with $c = \hbar = 1$ and a
metric signature of $(-1,1,1,1)$. The fundamental constants $G$ and
$\mu_0$ will be kept in the equations for better readability.

\section{Energy-momentum tensor in presence of a magnetic
  field}\label{s:magtmunu}

Matter properties enter the star's structure equations via the
energy-momentum tensor as the source of the Einstein equations and for
hydrodynamic equilibrium. Without the coupling to the electromagnetic
field, in the general relativistic context, it is generally assumed to
have the form of a perfect fluid,
\begin{equation}
T^{\mu\nu} = (\varepsilon + p)\; u^\mu u^\nu + p\; g^{\mu\nu}~,
\label{eq:perfectfluid}
\end{equation}
where $\varepsilon$ denotes the (matter) energy density, $p$ the
pressure, and $u^\mu$ the fluid four-velocity. The EoS then relates
pressure and energy density to the thermodynamically relevant
parameters, chosen following the system's equilibrium conditions. For
a cold neutron star in beta equilibrium without magnetic field, the
EoS is a function of only one parameter, often chosen to be baryon
number density, $p(n_B), \varepsilon(n_B)$. Another equivalent choice
is baryon chemical potential, which is at zero temperature equal to
the enthalpy per baryon $h$, i.e. $p(h), \varepsilon(h)$, adopted
here, following~\cite{Bonazzola1993, Bocquet}.

The aim of this section is to generalise the expression for the
energy-momentum tensor, Eq.~(\ref{eq:perfectfluid}), to the case of a
non-vanishing electromagnetic field taking into account the interaction of the
electromagnetic field with matter. Such an expression can be obtained by
taking the thermodynamic average of a microscopic energy-momentum tensor, see
next subsection. Neutron star matter is essentially composed of fermions, be
it hadronic (e.g. nucleons) or quark matter.  Hence, for this study we will
consider only the case of fermions and write down a general formalism for
fermions in an electromagnetic field. For the sake of clarity, we will thereby
neglect throughout the derivations any interaction apart from that with the
electromagnetic field. Additional interaction terms among the particles can
be added straightforwardly, see the next subsection ~\ref{ss:EoS},
where we present the model used for the numerical applications.

\subsection{Thermodynamic average of the microscopic 
 energy-momentum tensor}\label{ss:thermo_average}

Our starting point will be the microscopic energy-momentum tensor
obtained from the system's Lagrangian. The Lagrangian density of a
fermion system in the presence of a magnetic field can be written as
\begin{equation}
{\mathcal L} = - \bar{\psi}(x)(D_\mu \gamma^\mu + m)\psi(x) -\frac{1}{4 \mu_0}
F_{\mu\nu} F^{\mu\nu}~, 
\label{eq:qed}
\end{equation}
where $D_\mu = \partial_\mu - i \; q\; A_\mu$, $q$ denotes the charge of the
particles, and
\begin{equation}
F^{\mu\nu} = \partial^\mu A^\nu - \partial^\nu A^\mu ~, \label{e:def_fmunu}
\end{equation}
is the field strength tensor
of the electromagnetic field.  For the scales relevant the metric can
be assumed as (locally) flat, i.e. the Minkowski metric.

There are different ways to obtain the Einstein-Hilbert
energy-momentum tensor appearing as source of the Einstein
equations. It is defined via the requirement that the action,
\begin{equation}
S = \int {\mathcal L}\; \sqrt{-g} \, d^4x
\end{equation}
be invariant with respect to variations of the metric. This leads to
\begin{equation}
T^{\mu\nu} = \frac{-2}{\sqrt{-g}} \frac{\delta}{\delta g_{\mu\nu}}
(\sqrt{-g} {\mathcal L})~.
\end{equation}
\cite{Ferrer2010} use directly this definition to derive an expression
for $T^{\mu\nu}$. Since, however, it is not obvious to define fermion
fields within a non-flat metric, we will use here another strategy. In
flat space, following Noether's theorem, a divergence free
energy-momentum tensor can be derived from the invariance of the
Lagrangian with respect to translations in space and time.  It is,
however, neither symmetric nor gauge invariant, so that it is clearly
not suitable as source of the Einstein equations. However,
\cite{Belinfante} and \cite{Rosenfeld} have shown that it can be
written in a symmetrised gauge-invariant form by adding the divergence
free Belinfante-Rosenfeld correction term, see~\cite{Weinberg}. In
flat space, the Belinfante-Rosenfeld tensor is equivalent to the
Einstein-Hilbert energy-momentum tensor. In our case of a fermion
field coupled to an electromagnetic field it is given by
\begin{equation}
T^{\mu\nu} = -\frac{1}{\mu_0} F^{\mu\alpha} F_{\alpha}^{\;\nu} +
\frac{1}{2}\bar{\psi}(\gamma^\mu D^\nu + \gamma^\nu D^\mu) \psi + g^{\mu\nu}
{\mathcal L}~. 
\label{eq:tmunumicro}
\end{equation}
The first term represents the well-known contribution of the
electromagnetic field and the second term, arising from the fermion
field, agrees with Eq.~(36) in \cite{Ferrer2010} showing that indeed both
ways to evaluate the energy-momentum tensor are equivalent.

Since we are interested in studying the structure of a star on
macroscopic length scales, we need to calculate the thermodynamic
average of the microscopic energy-momentum tensor,
Eq.~(\ref{eq:tmunumicro}). It is assumed in the following derivations
that the electromagnetic fields are constant over the averaging
volume. The thermal average of $T^{\mu\nu}$ can then be written as,
see~\cite{Kapusta},
\begin{equation}   
  \langle T^{\mu\nu} \rangle = \frac{1}{\beta V} \frac{1}{Z} \int {\mathcal D}
  \psi {\mathcal D} \bar\psi \exp(\tilde S)  \,\int^\beta_0 d \lambda
  \int d^3 x \,T^{\mu\nu}~,
\end{equation}
where the partition function is given by
\begin{equation}
Z = \int {\mathcal D} \psi {\mathcal D} \bar\psi \exp(\tilde S) ~,
\end{equation}
and the action is
\begin{equation}
\tilde S = \int_0^\beta d\lambda \int d^3 x ({\mathcal L}(\lambda, x^i) -
\mu \hat{n})~.
\end{equation}
$\beta = 1/T$ is the inverse temperature, $\lambda = i x^0$, and the
term $\mu \hat{n}$ has to be introduced in grand canonical treatment
to guarantee average particle number conservation. The number density
operator is $\hat{n} = -i \bar{\psi} \gamma^0\hat{Q} \psi$, where the
operator $\hat{Q}$ associates the number density of the particle
species $a$ with its charge $Q_{a}$.  $\mu$ represents the associated
chemical potential.

The thermal average of the energy-momentum tensor is then given by
(see Appendix \ref{a:App1} for details of the calculations)
\begin{eqnarray}
\langle T^{\mu\nu}\rangle  &=& (\varepsilon + p)\; u^\mu u^\nu + p
  \; g^{\mu\nu} \nonumber \\ && + \frac{1}{2} (F^\nu_{\;\tau} M^{\tau\mu} +
F^{\mu}_{\;\tau} M^{\tau\nu} ) \nonumber \\ && 
 - \frac{1}{\mu_0} (F^{\mu\alpha} F_{\alpha}^{\,\nu} + \frac{g^{\mu\nu}}
{4} F_{\alpha\beta} F^{\alpha\beta}) ~.
\label{eq:tmunu}
\end{eqnarray}
The first two terms on the right hand side can be identified as the
pure (perfect fluid) fermionic contribution, followed by the
magnetisation term and finally the usual electromagnetic field
contributions to the energy-momentum tensor.  The magnetisation tensor
$M_{\mu \nu}$ is thereby defined as usual as the derivative of the
grand canonical potential with respect to the electromagnetic field
tensor, see Eq.~(\ref{eq:mtaudef}) and Eq.~(\ref{eq:defmmunu}). The
same form for the energy-momentum tensor\footnote{Note, however, the
  different metric convention.} has been given in the context of
special relativistic hydrodynamics in \cite{Huang} and for the case of
a perfect fluid + the electromagnetic field the above expression
agrees with \cite{Bonazzola1993}. From now on we will drop the
brackets indicating the thermal average for better readability.

In the fluid rest frame (FRF), assuming a perfect conductor, the
electric field vanishes and only the magnetic field $b_\mu$ is
nonzero. The electromagnetic field tensor can then be expressed in
terms of $b_\mu$ as~(\cite{Gourgoulhon})
\begin{equation}
F_{\mu\nu} = \epsilon_{\alpha\beta\mu\nu} u^{\beta} b^{\alpha}
\label{eq:deffmunu} 
\end{equation}
with the Levi-Civita tensor $\epsilon$, associated here with the
Minkowski metric. The above expression, Eq.~(\ref{eq:deffmunu}), is,
however, more general and can be employed with any metric.  If we
assume in addition, that the medium is isotropic and that the
magnetisation is parallel to the magnetic field, the magnetisation
tensor can be written as
\begin{equation}
M_{\mu\nu} = \epsilon_{\alpha\beta\mu\nu} u^{\beta} m^{\alpha} 
\label{eq:defmmunu}
\end{equation}
with the magnetisation four-vector 
\begin{equation}
m_\mu = \frac{x}{\mu_0} b_\mu~.
\label{e:def_x}
\end{equation}
As we shall see, the dependence of the different equations on the
magnetisation can now be reduced to a dependence on the scalar
quantity $x$, which can conveniently be computed in the FRF. First,
the energy-momentum tensor can be rewritten in the following way
\begin{eqnarray}
T^{\mu\nu}& =& (\varepsilon + p) \; u^\mu u^\nu + p\;
  g^{\mu\nu} \nonumber \\ && 
 + \frac{1}{\mu_0} \left( - b^\mu b^\nu + (b\cdot b) u^\mu u^\nu + \frac{1}{2}
   g^{\mu\nu} (b \cdot b) \right) \nonumber \\ &&
 + \frac{x}{\mu_0}  \left( b^\mu b^\nu - (b\cdot b)( u^\mu u^\nu + 
   g^{\mu\nu})\right) ~.
\end{eqnarray}
It is obvious that for a magnetic field pointing in $z$-direction this
expression reduces to the well-known form with magnetisation, see
e.g.~ \cite{Ferrer2010}. Neglecting the effect of magnetisation,
i.e. taking $x = 0$, it agrees with the standard MHD form, see
e.g.~\cite{Gourgoulhon}.

As already pointed out \textit{e.g.} by \cite{Potekhin}, there has
been some confusion in the literature about pressure anisotropy in the
presence of a magnetic field. From the above derivations it is clear
that the magnetic field does not induce any anisotropy to the matter
pressure defined thermodynamically as a derivative of the partition
function. It transforms as a scalar. The energy-momentum tensor,
however, shows anisotropies. If the spatial elements of the FRF
energy-momentum tensor are interpreted as pressures, then there is a
difference induced by the orientation of the magnetic field. Often the
different elements are called perpendicular and parallel pressures,
but they do not correspond to the thermodynamic pressure. Let us
stress that this anisotropy of the energy-momentum tensor does not
arise only from the magnetic field dependence of the EoS and the
magnetisation contribution, but that it is inherent already to the
purely electromagnetic energy-momentum tensor.

\cite{Blandford} (see \cite{Potekhin}, too) claim in addition that the
magnetisation contribution to the energy-momentum tensor is cancelled
by the Lorentz force associated with magnetisation. We shall see in
Section~\ref{ss:equil} that upon deriving the hydrodynamic equations
of motion for the system, this is indeed the case and that the
system's equilibrium depends only on $p, \varepsilon$ and the
electromagnetic field. We prefer, however, to keep the energy-momentum
tensor in its natural form, Eq.~(\ref{eq:tmunu}), including the
magnetisation, and add the Lorentz force to the equilibrium equations
via Maxwell equations, since we think that the physical origin of the
different contributions is presented in a clearer way.

\subsection{Equation of state}\label{ss:EoS} 

The evaluation of the matter pressure and energy density (EoS) for
different models of neutron star matter in the presence of a magnetic
field can be found in many papers in the (recent) literature, see
e.g.~\cite{Noronha, Rabhi2008, Ferrer2010, Rabhi2011, Strickland2012,
  Sinha}. Basically, charged particles become Landau
quantized~(\cite{Landau}) in the plane perpendicular to the magnetic
field. For our numerical applications, we will employ the quark model
in the MCFL phase to describe the neutron star interior. Let us now
briefly summarise the main characteristics of this model.

The effect of a strong magnetic field on quark matter was extensively
studied earlier by many authors, see e.g.~\cite{Gatto, Ferrer2013} and
references therein.  Here, we employ a simple massless three-flavor
MIT bag model, supplemented with a pairing interaction of NJL-type to
include the possibility of colour superconductivity in the
colour-flavor locked state similar to the model in \cite{Noronha,
  Paulucci},
\begin{equation}
{\mathcal L}_{\mathit{pairing}} =- \frac{G_P}{4} \sum_{\eta = 1}^3 (\bar{\psi}
P_{\eta} C \bar{\psi}^T) (\psi^T C \bar{P}_{\eta} \psi)~, 
\end{equation}
where $C = i \gamma^2 \gamma^0$ is the charge conjugation matrix. The
quark spinors $\psi^\alpha_a$ carry colour $a = (1,2,3)$ and flavour
$\alpha = (s,d,u)$ indices. $\bar{P}_{\eta} = \gamma^0 P_\eta^\dagger
\gamma^0$, and the considered pairing matrix is given by
$(P_\eta)^{ab}_{\alpha\beta} = i \gamma^5 \epsilon^{a b \eta}
\epsilon_{\alpha\beta\eta}$, i.e. we only take pairing in
antisymmetric channels into account. Following the same scheme as in
\cite{Noronha}, we computed the EoS of quark matter in the MCFL phase,
using $G_P = 5.15\ \mathrm{GeV}^{-2}, \Lambda = 1$ GeV and a bag
constant $B_{\mathit{bag}} = 60 $ MeV/fm$^3$.

\begin{figure}
  \begin{center}
      \includegraphics[width=.4\textwidth]{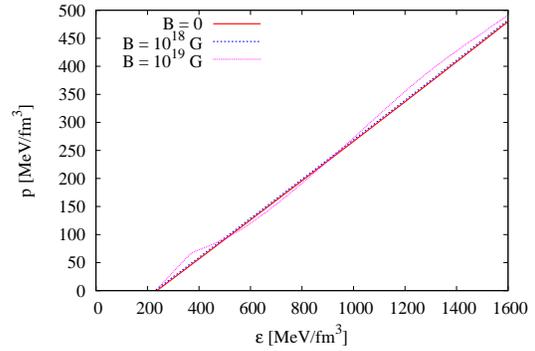}
    \caption{EoS of quark matter in MCFL phase for different magnetic fields}
    \label{fig:eosmcfl}
  \end{center}
\end{figure}

The EoS for different constant magnetic field values is displayed in
Fig.~(\ref{fig:eosmcfl}).  The effect of the magnetic field starts to
become significant only at very large fields ($B \gtrsim 10^{19}$
G). The observed oscillations are due to the de Haas-van Alphen
oscillations, pointed out already in \cite{Noronha}.

\begin{figure}
  \begin{center}
      \includegraphics[width = .4\textwidth]{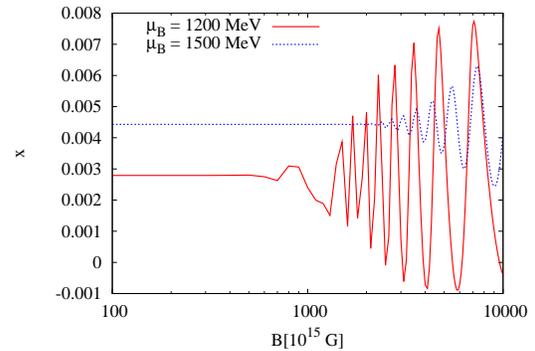}
    \caption{Magnetisation divided by magnetic field as a function of magnetic
    field strength in the MCFL phase. }
    \label{fig:pdiffb}
  \end{center}
\end{figure}

The quantity $x$, corresponding in the FRF to the magnetisation
divided by the magnetic field is shown in Fig.~(\ref{fig:pdiffb}) for
two different values of baryon number chemical potential. The values
are in agreement with Fig.~2 of~\cite{Noronha}. It is obvious that the
magnetisation in this model is too small for reasonable values of the
magnetic field reachable in magnetars, to induce any considerable
change in the neutron star structure. This will be confirmed by the
numerical results in Section~\ref{s:resu}.

\section{Global models in the stationary and axisymmetric
  case}\label{s:global}

In order to explore the effects of the inclusion of the magnetic field
onto the neutron star structure and properties, we have numerically
computed, within the framework of general relativity, complete models
of rotating neutron stars endowed with a magnetic field. In this
section, we present the physical model built to obtain the global
models and the equations that are solved. Note that, in this section,
Latin letters $i, j, \dots$ are used for spatial indices only, whereas
Greek ones $\alpha, \mu, \dots$ denote the spacetime indices.

Within the theory of general relativity for the gravitational field,
we follow the approach by \cite{Bonazzola1993} and make the assumption
of a stationary, axisymmetric spacetime, in which the matter content
(the energy-momentum tensor) fulfils the {\em circularity
  condition}. The line element expressed in spherical-like coordinates
then reads:
\begin{eqnarray}
  {\rm d}s^2 &=& -N^2\, {\rm d}t^2 + B^2r^2\sin^2\theta \left({\rm d}\varphi -
    N^\varphi\, {\rm d}t \right)^2 \nonumber\\
  &&+ A^2\left( {\rm d}r^2 + r^2\, {\rm d}\theta^2\right),\label{e:def_metric}
\end{eqnarray}
where $N, N^\varphi, A$ and $B$ are functions of coordinates $(r,
\theta)$. 

\subsection{Maxwell equations}\label{ss:Maxwell}

In the same way as in~\cite{Bocquet} we consider here that the
electromagnetic field originates from free currents, noted hereafter
simply $j^\sigma$, which are \textit{a priori} independent from the
movements of inert mass (with 4-velocity $u^\mu$). This is a limiting
assumption in our model, and one should in principle use a microscopic
model to derive a distribution for the free currents, too. However,
such a model would require a multi-fluid approach to model the
movements of free protons and electrons, and we leave it for a future
study.

Under the symmetries defined in our model (see beginning of
Section~\ref{ss:GR}) the four-potential $A_\mu$, entering in the
definition of the electromagnetic field tensor $F^{\mu\nu}$ through
Eq.~(\ref{e:def_fmunu}), can induce either purely poloidal or purely
toroidal magnetic fields (\cite{Frieben}). Here, we chose a purely
poloidal configuration, meaning in particular that the four-potential
has vanishing components $A_r=A_\theta = 0$. The electric and magnetic
fields measured by the Eulerian observer (whose four-velocity is
$n^\mu$) are then defined as $E_\mu = F_{\mu\nu}\, n^\nu$ and $B_\mu =
-\frac{1}{2} \epsilon_{\mu\nu\alpha\beta}\, n^\nu\, F^{\alpha\beta}$,
with $\epsilon_{\mu\nu\alpha\beta}$ the Levi-Civita tensor associated
with the metric~(\ref{e:def_metric}). The non-zero components read:
\begin{subequations}
  \begin{eqnarray}
    E_r & = & \frac{1}{N} \left( \df{A_t}{r} + N^\varphi
      \df{A_\varphi}{r} \right)\label{e:def_Er}\\
    E_\theta & = & \frac{1}{N} \left( \df{A_t}{\theta} + N^\varphi
      \df{A_\varphi}{\theta} \right) \label{e:def_Et}\\ 
    B_r & = &
    \frac{1}{Br^2\sin\theta}\df{A_\varphi}{\theta} \label{e:def_Br}\\
    B_\theta & = & - \frac{1}{B \sin \theta} \df{A_\varphi}{r} \label{e:def_Bt}
  \end{eqnarray}
\end{subequations}

The homogeneous Maxwell equation $F_{[\mu\nu;\lambda]} = 0$
(Faraday-Gauss) is automatically fulfilled, when taking the form in
Eq.~(\ref{e:def_fmunu}) for the tensor $F^{\mu\nu}$. The inhomogeneous
Maxwell equation (Gauss-Amp\`ere) in presence of external magnetic
field ($\nabla_\mu$ is the covariant derivative associated with the
metric~(\ref{e:def_metric})),
\begin{equation}
  \frac{1}{\mu_0} \nabla_\mu F^{\nu\mu} = j_{\ \mathit{free}}^\nu +
  \nabla_\mu M^{\nu\mu}~,
\end{equation}  
can then be transformed to give
\begin{equation}
  \nabla_\mu F^{\sigma\mu} = \frac{1}{1-x} (\mu_0 j_{\ \mathit{free}}^\sigma +
  F^{\sigma\mu}\nabla_\mu x)~.
\label{eq:gauss-ampere}
\end{equation}  
This equation can be expressed in terms of the two non-vanishing
components of $A^\mu$, with the Maxwell-Gauss equation
\begin{eqnarray}
  \Delta_3 A_t &=& \frac{1}{x-1} \times \left[ \mu_0 A^2 \left(g_{tt}
      j_{\ \mathit{free}}^t + g_{t\varphi} j_{\ \mathit{free}}^\varphi
    \right)  + \partial A_t\, \partial x \right] \nonumber \\ 
  && - \frac{B^2}{N^2} N^\varphi r^2
  \sin^2\theta \partial A_t\, \partial N^\varphi \nonumber\\
  && - \left( 1 + \frac{B^2}{N^2} r^2 \sin^2\theta \left( N^\varphi
    \right)^2 \right) \partial A_\varphi\, \partial N^\varphi \nonumber \\
  && - \left( \partial A_t + 2N^\varphi \partial A_\varphi
  \right) \partial \left( \beta - \nu \right) \nonumber \\
  && - 2 \frac{N^\varphi}{r} \left( \df{A_\varphi}{r} + \frac{1}{r\tan
      \theta} \df{A_\varphi}{\theta} \right)~, \label{e:maxwell_gauss} 
\end{eqnarray}
and the Maxwell-Amp\`ere equation
\begin{eqnarray}
  \tilde{\Delta}_3 \left( \frac{A_\varphi}{r\sin\theta} \right) &=&
  \frac{1}{x-1} \times \left[ \mu_0 A^2 B^2 \left( j_{\ \mathit{free}}^\varphi -
      N^\varphi j_{\ \mathit{free}}^t \right) r \sin\theta \right. \nonumber \\
  && \left. + \frac{1}{r\sin\theta} \partial A_\varphi\, \partial x
  \right] \nonumber \\
  && + \frac{B^2}{N^2} r \sin\theta\, \partial N^\varphi \left( \partial
    A_t + N^\varphi \partial A_\varphi \right) \nonumber \\
  && + \frac{1}{r\sin\theta} \partial A_\varphi\, \partial \left( \beta -
    \nu \right)~, \label{e:maxwell_ampere}
\end{eqnarray}
with the following notations:
\begin{eqnarray*}
  & & \nu = \ln N, \qquad \alpha = \ln A, \qquad \beta = \ln B, \\
  & & \Delta_2 = \df{^2}{r^2} + \frac{1}{r}\df{}{r} + \frac{1}{r^2}
  \df{^2}{\theta^2} \\
  & & \Delta_3 = \df{^2}{r^2} + \frac{2}{r} \df{}{r} +
  \frac{1}{r^2}\df{^2}{\theta^2} + \frac{1}{r^2\tan \theta}
  \df{}{\theta}\\
  & & \tilde{\Delta}_3 = \Delta_3 - \frac{1}{r^2\sin^2 \theta}\\
  & & \partial a\, \partial b = \df{a}{r}\df{b}{r} + \frac{1}{r^2}
  \df{a}{\theta}\df{b}{\theta}. 
\end{eqnarray*}
In the case without magnetisation, ($x=0$), Eqs.~(6) and (7) of \cite{Bocquet}
are recovered.

\subsection{Einstein equations and energy-momentum
  tensor}\label{ss:GR}

Under the present assumptions of a stationary, axisymmetric spacetime,
the Einstein equations result in a set of four elliptic partial
differential equations for the metric potentials defined in
Eq.~(\ref{e:def_metric}):
\begin{subequations}
  \label{e:grav}
  \begin{eqnarray}
    & &\Delta_3 \nu = 4\pi G A^2 \left(E + S^i_{\ i}\right) +
    \frac{B^2 r^2 \sin^2 \theta}{2N^2}(\partial N^\varphi)^2 \nonumber \\
    & & \qquad \qquad - \partial\nu\, \partial(\nu+\beta) \label{e:grav_nu}\\
    & & \tilde{\Delta}_3 \left( N^\varphi r  \sin \theta \right) =
    -16\pi G \frac{NA^2}{B} \frac{J_\varphi}{r \sin \theta}
    \nonumber \\
    & & \qquad \qquad \qquad \qquad - r\sin\theta \, \partial
    N^\varphi\, \partial(3\beta - \nu)\label{e:grav_nphi}\\
    & & \Delta_2 \left[ \left(NB - 1 \right)r\sin\theta \right]
    \nonumber\\  
    & & =
    8\pi G N A^2 B r\sin \theta \left( S^r_{\ r} +
      S^\theta_{\ \theta} \right) \label{e:grav_G}\\
    & & \Delta_2 \left( \nu + \alpha \right) = 8\pi G A^2 S^\varphi_{\
      \varphi} + \frac{3B^2 r^2 \sin^2 \theta}{4N^2} \left( \partial N^\varphi
    \right)^2 \nonumber \\
    & & \qquad \qquad \qquad - \left( \partial \nu \right)^2, \label{e:grav_dzeta}
  \end{eqnarray}
\end{subequations}
with the same notations as those introduced in
Eqs.~(\ref{e:maxwell_gauss},\ref{e:maxwell_ampere}). 

Finally, $E, J_i, S^i_{\ j}$ are quantities obtained from the
so-called 3+1 decomposition of the energy-momentum tensor (for
definitions, see \textit{e.g.}~\cite{Gourgoulhon}). In our case of
Eq.~(\ref{eq:tmunu}) describing a perfect fluid endowed with a
magnetic field, including magnetisation effects, they can be written
in axisymmetric stationary symmetries as:
\begin{subequations}
  \begin{eqnarray}
    E &=& \Gamma^2 \left( \varepsilon + p \right) - p \nonumber \\
    && + \frac{1}{2\mu_0} \left[ (1 + 2x) E^iE_i + B^i B_i
    \right], \label{e:Etot} \\ 
    J_\varphi &=& \Gamma^2 \left( \varepsilon + p \right)U \nonumber\\
    && + \frac{1}{\mu_0} \left[ A^2 \left(B^r E^\theta - E^r B^\theta
      \right) + x B^i B_i U \right], \label{e:Jphitot}\\
    S^r_{\ r} &=& p + \frac{1}{2\mu_0} \left( E^\theta E_\theta - E^r
      E_r + B^\theta B_\theta - B^r B_r \right. \nonumber \\
    && \left. + \frac{2x}{\Gamma^2} B^\theta B_\theta \right), \label{e:Srrtot}\\
    S^\theta_{\ \theta} &=& p + \frac{1}{2\mu_0} \left( E^r
      E_r - E^\theta E_\theta + B^r B_r - B^\theta B_\theta \right. \nonumber \\
    && \left. + \frac{2x}{\Gamma^2} B^r B_r \right), \label{e:Stttot} \\
    S^\varphi_{\ \varphi} &=& p + \Gamma^2 \left( \varepsilon + p
    \right)U^2 + \frac{1}{2\mu_0} \left[ E^i E_i + B^i B_i
    \right. \nonumber \\
    && \left. + \frac{2x}{\Gamma^2} \left(1 + \Gamma^2 U^2\right) B^i
      B_i \right], \label{eSpptot} 
  \end{eqnarray}
  \label{e:3p1_Tmunu}
\end{subequations}
all other components of $J_i$ and $S^i_{\ j}$ being zero. $x$ is the
magnetisation, defined by Eq.(\ref{e:def_x}) and $U$ is the physical
fluid velocity in the $\varphi$ direction, as measured by the Eulerian
observer; it is given by
\begin{equation}
  \label{e:def_U}
  U = \frac{Br\sin\theta}{N} \left( \Omega - N^\varphi \right),
\end{equation}
with $\Omega = u^\varphi / u^t$ being the fluid coordinate angular
velocity (gauge independent). The electric ($E^i$) and magnetic
($B^i$) fields have been defined in Section~\ref{ss:Maxwell}.

\subsection{Magnetostatic equilibrium}\label{ss:equil}

The equations for magnetostatic equilibrium can be derived from the
conservation of energy and momentum, expressed as vanishing divergence
of the energy-momentum tensor:
\begin{equation}
\nabla_\mu T^{\mu\nu} = 0.
\end{equation}

This can be detailed as :
\begin{equation}
  \nabla_\mu T^{\alpha\beta} = \nabla_\alpha T_f^{\alpha\beta}
  -F^{\beta\nu} j^{\mathit\ free}_{\nu}  - \frac{x}{2\mu_0}
  F_{\sigma\tau} \nabla^\beta F^{\sigma\tau}~, 
\end{equation}
where $T_f^{\alpha\beta}$ represents the perfect-fluid contribution to
the energy-momentum tensor; one can recognise the usual Lorentz force
term, too, arising from free currents. In the absence of
magnetisation, the expression is the same as in \cite{Bonazzola1993}.

As in \cite{Bocquet}, in the case of rigid rotation ($\Omega$ constant
across the star), a first integral of the following expression is
sought
\begin{eqnarray}
  (\varepsilon + p) \left ( \frac{1}{\varepsilon + p} \frac{\partial
      p}{\partial x^i} + \frac{\partial \nu}{\partial x^i} -
    \frac{\partial \ln \Gamma}{\partial x^i}\right)  
  -  F_{i \rho} j^\rho_{\ \mathit{free}} && \nonumber \\
  - \frac{x}{2\mu_0}  F_{\mu\nu}
  \nabla_i F^{\mu\nu} &=& 0~.
\label{eq:first-integral}
\end{eqnarray}

In order to obtain this first integral, one introduces the enthalpy
per baryon and its derivatives. It can be shown that, even in the
presence of the magnetic field, the logarithm of the enthalpy per
baryon represents again a first integral of the fluid equations. To
that end, let us first note that for the neutron star case with a
magnetic field in beta-equilibrium and at zero temperature, the
enthalpy is a function of both baryon density and magnetic field
\begin{equation}
 h = h(n_b, b) = \frac{\varepsilon + p}{n_b} = \mu_b~.\label{e:def_h}
\end{equation} 
Hence we have
\begin{equation}
  \frac{\partial \ln h} {\partial x^i} = \frac{1}{h} \left(\left. \frac{\partial
        h}{\partial n_b} \right |_b \frac{\partial n_b} {\partial x^i} +
    \left. \frac{\partial
        h}{\partial b} \right |_{n_b} \frac{\partial b} {\partial x^i}\right)~.
\end{equation}
In addition, the following thermodynamic relations are valid under the
present assumptions
\begin{eqnarray}
  \left.\frac{\partial h}{\partial n_b}\right |_b &=& \frac{1}{n_b}
  \frac{\partial p}{\partial n_b} \\
  \left. \frac{\partial p}{\partial b}\right|_{\mu_b} &=& m ( =
  \sqrt{m^\mu m_\mu}) = - 
  \left. \frac{\partial
      \varepsilon}{\partial b} \right|_{n_b}~.
\end{eqnarray}
And we obtain for the derivative of the logarithm of the enthalpy
\begin{eqnarray}
  \frac{\partial \ln h} {\partial x^i} &=& \frac{1}{\varepsilon + p}
  \left[\left. \frac{\partial p}{\partial n_b}\right |_b
    \frac{\partial n_b} {\partial x^i} + \left(\left. \frac{\partial
          p}{\partial b} \right |_{n_b} - m \right) \frac{\partial b} 
    {\partial x^i}\right] \nonumber \\ &=& \frac{1}{\varepsilon + p}
  \left( \frac{\partial p}{\partial x^i} - m \frac{\partial
      b}{\partial x^i} \right)~.\label{e:dlnh}
\end{eqnarray}

The second term in Eq.~(\ref{eq:first-integral}) -- $F_{i \rho}
j^\rho_{\ \mathit{free}}$ -- is treated as in \cite{Bonazzola1993} and
we assume that \textit{i)} matter is a perfect conductor ($A_t =
-\Omega A_\varphi$ inside the star); \textit{ii)} it is possible to
relate the components of the electric current to the electromagnetic
potential $A_\varphi$, through an arbitrary function $f$, called the
{\em current function\/}:
\begin{equation}
  \label{e:current_function}
  j^\varphi - \Omega j^t = (\varepsilon + p) f\left( A_\varphi
  \right). 
\end{equation}
Under these two assumptions, the Lorentz force term becomes
\begin{equation}
  \label{e:Lorentz_force}
  F_{i \rho} j^\rho_{\ \mathit{free}} = \left( j^\varphi - \Omega j^t
  \right) \df{A_\varphi}{x^i} = - \left(\varepsilon + p \right) \df{M}{x^i},
\end{equation}
with
\begin{equation}
  \label{e:def_M}
  M(r, \theta) = - \int_0^{A_\varphi(r,\theta)} f(x) {\rm d}x.
\end{equation}

The last term can be written in terms of the magnetic field $b^\mu$ in
the FRF as (with $b^2 = b_\mu b^\mu$):
\begin{equation}
  \frac{x}{2\mu_0} F_{\mu\nu} \nabla_i F^{\mu\nu} =
  \frac{x}{\mu_0}\left( b_\mu \nabla_i b^\mu - b_\mu b^\mu u_\nu \nabla_i
    u^\nu \right) = b \nabla_i b = m \df{b}{x^i},
\end{equation} 
from the expression~(\ref{eq:deffmunu}), and the definition~(\ref{e:def_x}).

Thus, this last term cancels with its counterpart in
Eq.~(\ref{e:dlnh}) and the first integral (\ref{eq:first-integral})
keeps exactly the same form as without magnetisation:
\begin{equation}
  \label{e:1st_integral}
  \ln h(r, \theta) + \nu(r, \theta) - \ln \Gamma(r, \theta) + M(r,
  \theta) = {\rm const}.
\end{equation}

\subsection{Numerical resolution}\label{ss:numerics}

The equations have been solved with the library~\cite{Lorene}, using
spectral methods to solve Poisson-like partial differential equations
appearing in the Einstein-Maxwell system~(\ref{e:grav}),
(\ref{e:maxwell_gauss}) and (\ref{e:maxwell_ampere}). For more details
about these methods, see \textit{e.g.}  \cite{grandclement-09}. The
code follows the algorithm presented by \cite{Bocquet}, but with the
modification of the inclusion of new magnetisation terms,
\textit{i.e.} depending on the magnetisation $x$ defined in
Eq.~(\ref{e:def_x}), in these partial differential equations. However,
as it has been shown in Eq.~(\ref{e:1st_integral}) the expression for
the equilibrium of the fluid in the gravitational and magnetic fields
does not change.

The most important difference with \cite{Bocquet} comes from the use
of an EoS which gives all the needed variables: $p, \varepsilon, n_b,
x$; depending on two parameters (instead of one): the enthalpy
$h$~(\ref{e:def_h}) and the magnetic field amplitude in the FRF $b =
\sqrt{b_\mu b^\mu}$. These quantities are first computed and stored on
a table once for all. This is then read by the code computing the
equilibrium global models, and a bi-dimensional interpolation using
Hermite polynomials is used, following the method described by
\cite{swesty-96}, to ensure thermodynamic consistency of the
interpolated quantities ($p(h, b), \varepsilon(h, b), n_b(h, b)$ and
$x(h, b)$).
\begin{figure*}
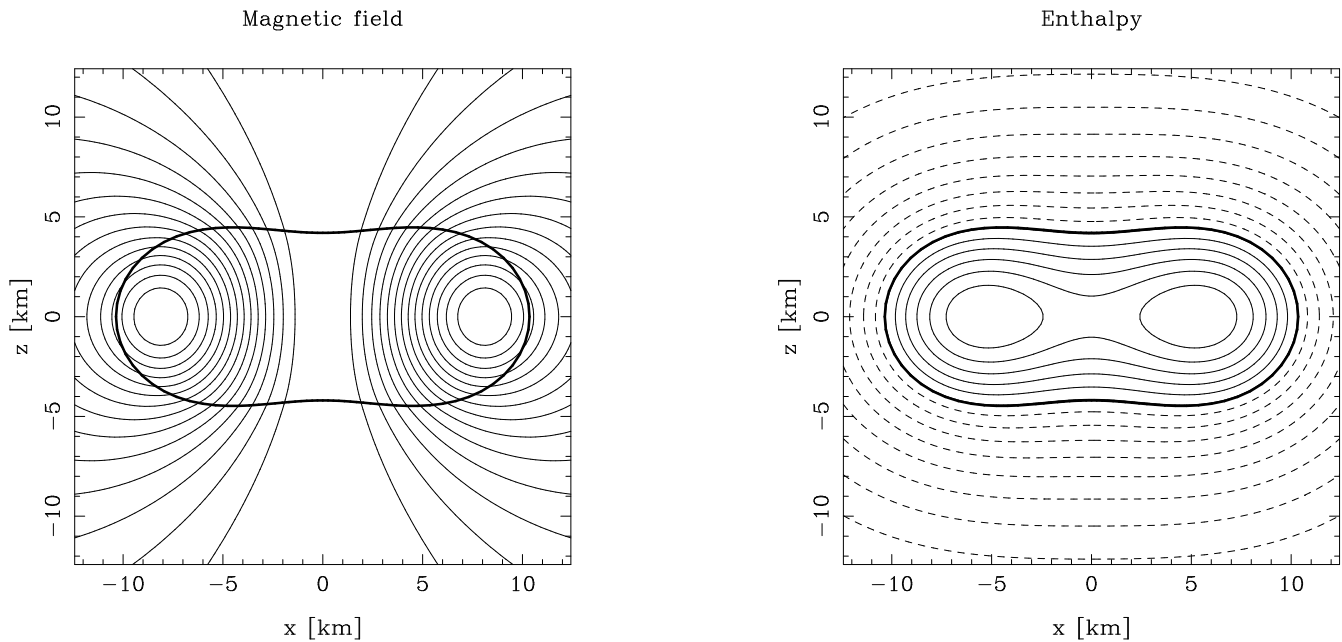

  \begin{center}
    \includegraphics[scale=.45,angle=270]{maxB.ps}\hfill
    \includegraphics[scale=.45,angle=270]{ent_maxB.ps}
  \end{center}
  \caption{Magnetic field lines (left) and enthalpy isocontours
    (right) in the meridional plane $(x, z)$, for the static star
    configuration, with a gravitational mass of $2.22 M_\odot$ and a
    polar magnetic field of $8.16 \times 10^{17}$~G. The stellar
    surface is depicted by the bold line. In the right figure, solid
    lines represent positive enthalpy isocontours, dashed lines
    negative ones (no matter).}
  \label{fig:maxB_slices}
\end{figure*}

The free physical parameters entering our model are: the EoS, the
current function $f$~(\ref{e:current_function}), the rotation
frequency $\Omega$ and the logarithm of the central enthalpy $H_c =
\log(h(r=0))$. Once the equilibrium configuration has been computed,
global quantities are obtained either from integration over the star's
volume (\textit{e.g.} baryonic mass $M_B$) or from the asymptotic
behaviour of the gravitational field (\textit{e.g.} gravitational mass
$M_G$) and of the electromagnetic field (\textit{e.g.} magnetic moment
$\mathcal{M}$). Detailed definitions and formulae can be found
\cite{Bonazzola1993} and \cite{Bocquet}.

\section{Results and Discussion}\label{s:resu}

We computed models of fully relativistic neutron stars with a poloidal
magnetic field, employing the EoS described in Sec.(\ref{ss:EoS}), and
a constant current function~(\ref{e:current_function}) $f(x)= f_0$. As
shown in \cite{Bocquet}, the choice of other current functions for $f$
would not alter the conclusions. Varying $f_0$ allowed us to vary the
intensity of the magnetic field, as measured for instance by the value
of the radial component at the star's pole (polar magnetic field), or
by the magnetic moment. The variation of the central enthalpy has a
direct influence on the star's masses ($M_B$ and $M_G$), although they
depend on the rotation frequency and magnetic field strength, too. To
demonstrate pure magnetic field effects on the neutron star
configurations, we first computed static neutron stars.

The first point to emphasise is that, as it has already been
illustrated, e.g. in \cite{Bocquet, Cardall}, the stellar
configurations can strongly deviate from spherical symmetry due to the
anisotropy of the energy-momentum tensor in presence of a
non-vanishing electromagnetic field. As an example we show in
Fig.~(\ref{fig:maxB_slices}) the magnetic field lines and the enthalpy
profile in the $(r,\theta)$-plane for a configuration with a magnetic
moment of $3.25 \times 10^{32}$ A m$^2$ and a baryon mass of 2.56
$M_{\odot}$. These values correspond to a polar magnetic field of 8.16
$\times 10^{17}$ G and a gravitational mass of 2.22 $M_{\odot}$. The
asymmetric shape of the star due to the Lorentz forces exerted by the
electromagnetic field on the fluid is evident from the figures. Upon
increasing the magnetic field strength the star's shape becomes more
and more elongated, finally reaching a toroidal shape, see
\cite{Cardall}. However, our code is not able to treat this change of
topology and the configuration shown in Figs.~\ref{fig:maxB_slices}
represents the limit in terms of magnetic field strength, that can be
computed within our numerical framework. Therefore for this study, we
compute stellar configurations within this maximum
limit. Nevertheless, note that the polar magnetic field value is well
above any observed magnetic field in magnetars.

\begin{figure}
  \begin{center}
      \includegraphics[height=9cm,width=8cm,angle=270]{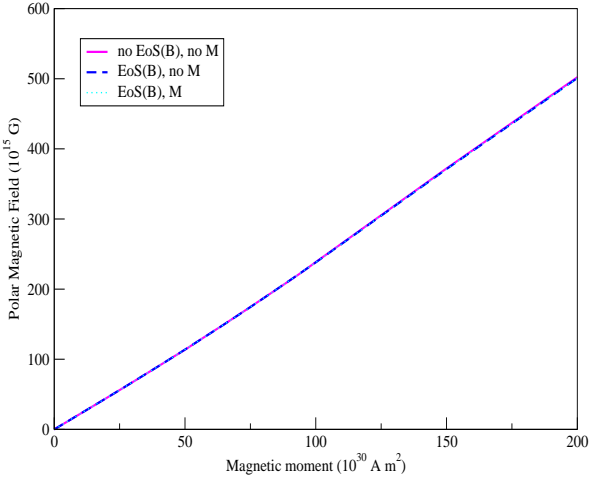}
      \caption{Polar magnetic field as a function of magnetic moment
        for constant current function and baryonic mass 1.6
        $M_{\odot}$ with and without magnetic field dependence and
        magnetisation (see text for details).}
    \label{fig:bpmm}
  \end{center}
\end{figure}

The determination of the maximum gravitational mass is usually
performed considering sequences of constant magnetic moment
$\mathcal{M}$ and increasing central enthalpy $H_c$ (see
\cite{Bocquet}). To be able to relate better with astrophysical
observations of magnetars, in Fig.~(\ref{fig:bpmm}) we plot the value
of the polar magnetic field corresponding to the values of the
magnetic moment for a neutron star having baryonic mass 1.6
$M_{\odot}$.
In this figure, three curves have been plotted, corresponding to three
types of configurations:
\begin{enumerate}
\item A full model as described in
Sect.~\ref{s:global}, denoted by EoS(B), M;
\item A model with magnetic field dependence of the EoS, but no
  inclusion of the magnetisation terms $x$ in the energy-momentum
  tensor -- setting $x=0$ in Eqs. (\ref{eq:gauss-ampere}) and (\ref{e:3p1_Tmunu}) -- denoted by EoS(B), no M;
\item A bare model where both these effects are excluded (no EoS(B), no M), which
is a case comparable with the study by \cite{Bocquet}.
\end{enumerate}
These settings shall be used later in this work, too. The polar
magnetic field increases linearly with the magnetic moment, and is
indistinguishable between the three cases discussed above, i.e. with
and without inclusion of magnetic field dependence of the EoS and
magnetisation. The relation between the magnetic moment and the polar
magnetic field changes only slightly depending on the baryon mass of
the star, the present figure can therefore be used as a guideline for
all the configurations shown within this work.

\begin{figure}
  \begin{center}
    \includegraphics[height = .53\textwidth,angle=270]{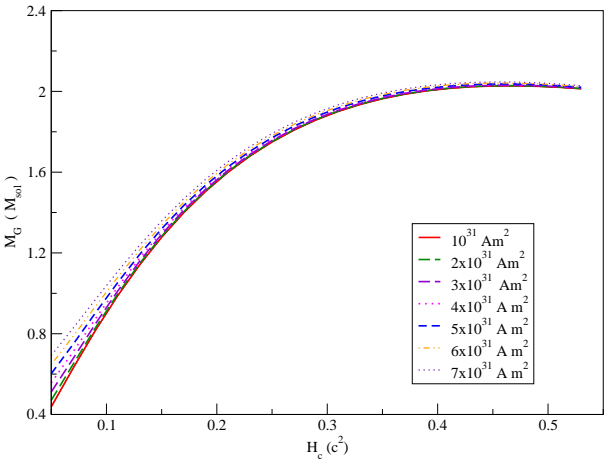}
    \caption{Gravitational mass as a function of central log-enthalpy
      $H_c$, along seven constant curves of magnetic dipole moment $\cal{M}$
      for non-rotating configurations.}
    \label{fig:hcmg}
  \end{center}
\end{figure}

We further studied the influence of using magnetic field dependent EoS
on the neutron star maximal mass, and we computed static
configurations determined by different values of central log-enthalpy
$H_c$ along constant curves of magnetic dipole moment $\cal{M}$, and
for each of them we plotted the gravitational masses in
Fig.~(\ref{fig:hcmg}). We then determined the maximum gravitational
mass ($M_G^{\rm max}$) corresponding to each values of magnetic moment
by parabolic interpolation.

\begin{figure*}
  \begin{center}
    \centerline{
      \includegraphics[height=9cm,width=8cm,angle=270]{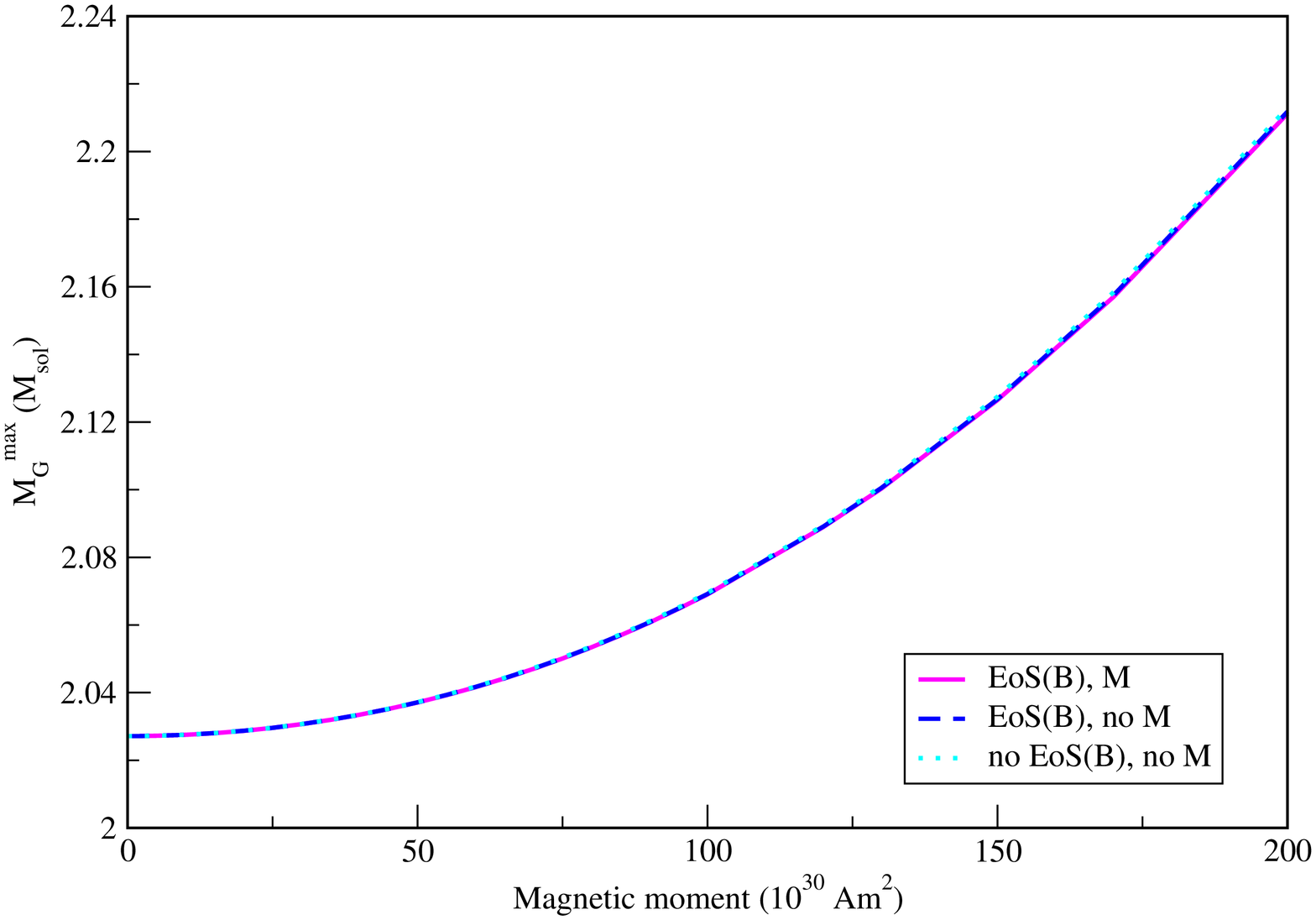}\hfill
      \includegraphics[height=9cm,width=8cm,angle=270]{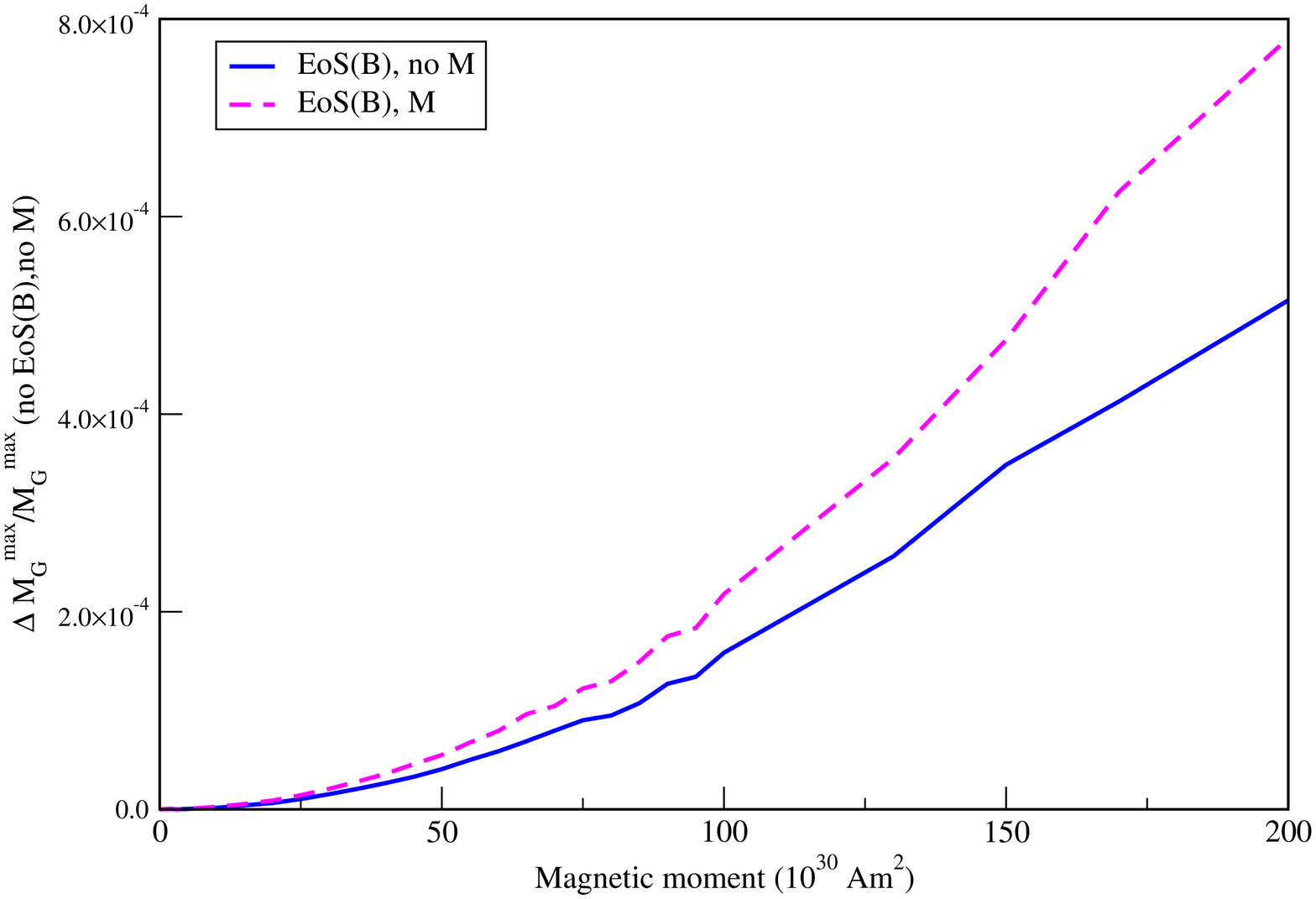}
    }
    \caption{Neutron star maximal mass (left panel) and relative
      difference in this mass among three models, as a function of
      magnetic moment. The three models correspond to the
      possibility or not of including of magnetisation term $x$ (``M''
      or ``no M''), and
      to the magnetic field dependence or not of the EoS (``EoS(B)'' or
      ``no EoS(B)'').}
    \label{fig:mmmgdiff}
  \end{center}
\end{figure*}

As noted before in \cite{Bocquet}, the maximal gravitational mass is
an increasing function of $\cal{M}$, and we have retrieved this result
in the left panel of Fig.~\ref{fig:mmmgdiff}.  It is evident that
there is very little difference on inclusion of the full model, with
respect to the one by \cite{Bocquet}, but in order to be more precise,
we plotted in the right panel of Fig.~(\ref{fig:mmmgdiff}) the
relative differences in the maximum gravitational masses as functions
of the magnetic moment, with and without the inclusion of the above
magnetic field effects compared to the case excluding these
effects. In this right panel, we see that even for very high magnetic
moments, corresponding to polar magnetic field much higher than those
observed in magnetars (see Fig.~\ref{fig:bpmm} for correspondence),
the relative difference in the maximal mass of magnetised neutron
stars is at most of the order $10^{-3}$ and therefore negligible
compared with uncertainties existing in the EoS models.

\begin{figure}
  \begin{center}
       \includegraphics[height=9cm,width=8cm,angle=270]{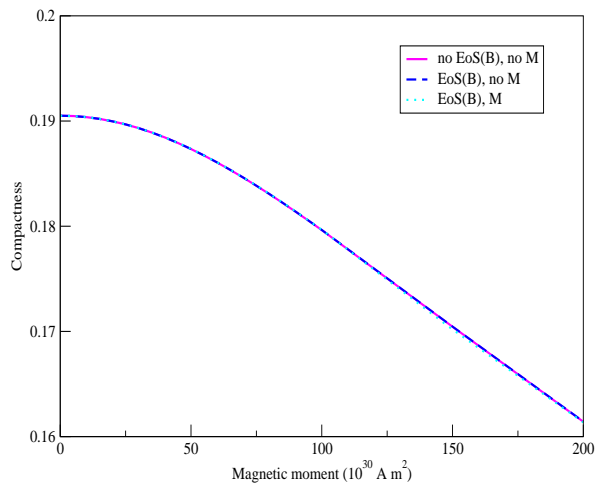}
       \caption{Compactness as a function of magnetic moment for
         neutron star with baryon mass 1.6~$M_{\odot}$ with and
         without magnetic field dependence and magnetisation (see
         Fig.\ref{fig:mmmgdiff}).}
    \label{fig:compact}
  \end{center}
\end{figure}

Another neutron star parameter of astrophysical interest is the
compactness $\mathcal{C}$, which is the dimensionless ratio of the
gravitational mass and radius
\begin{equation}
  \label{e:def_compactness}
  \mathcal{C} = \frac{GM_G}{R_{\rm circ} c^2},
\end{equation}
where $R_{\rm circ}$ is the circumferential equatorial radius (see
\cite{Bonazzola1993}).  We studied the behaviour of the compactness of
a neutron star of baryon mass 1.6 $M_{\odot}$ with magnetic moment, as
illustrated in the Fig.~(\ref{fig:compact}). The compactness was found
to decrease with increase in magnetic moment. This is understandable
from the centrifugal forces exerted by the Lorentz force on matter at
the centre, increasing with increasing magnetic moment, i.e. magnetic
field, see e.g. the discussion in \cite{Cardall}. Again the lines
corresponding to the cases with and without magnetisation or magnetic
field effects in the EoS are almost indistinguishable and the main
effect arises from the purely electromagnetic part already included in
\cite{Bocquet}.

As neutron stars are expected to be good sources of gravitational
radiation, we have looked at the influence of magnetisation and
magnetic field-dependent EoS on this mechanism. The scenario we
studied was the emission of gravitational waves when the magnetic
dipole is not aligned with the rotation axis, and magnetic field
deformation induces a time variation of the quadrupole moment. This
setting has previously been addressed by \cite{bonazzola-96}, who have
shown that, using the standard quadrupole formula to compute the
characteristics of gravitational waves, it was possible to split the
quadrupole moment into two parts: a first depending on the rotation,
and a second on the distortion of the star. Thus, it is possible to
study the amplitude of gravitational waves by looking at the
deformation of static (non-rotating) stars. We therefore considered a
sequence of non-rotating stars, with fixed baryon mass $M_B = 2
M_\odot$ and increasing magnetic moment. For this sequence, we
looked at the quadrupole moment $Q$, as defined in Eq.~(7) of
\cite{salgado-94}, as a function of the magnetic moment
$\mathcal{M}$. First, we recovered the behaviour $Q\sim
\mathcal{M}^2$, given in Eq.~(32) of~\cite{bonazzola-96}. Then, we
looked again at the relative difference in $Q$, between a sequence
using the full approach devised here, and a simpler one with the EoS
not depending on the magnetic field and no magnetisation
term. Contrary to the maximum mass case, this difference remains
almost independent of the magnetic moment, with a value $\sim
10^{-3}$. This shows that the gravitational wave emission properties
are very little sensitive to the use of a magnetic-field dependent
EoS.

\begin{figure}
  \begin{center}
        \includegraphics[height=9cm,width=8cm,angle=270]{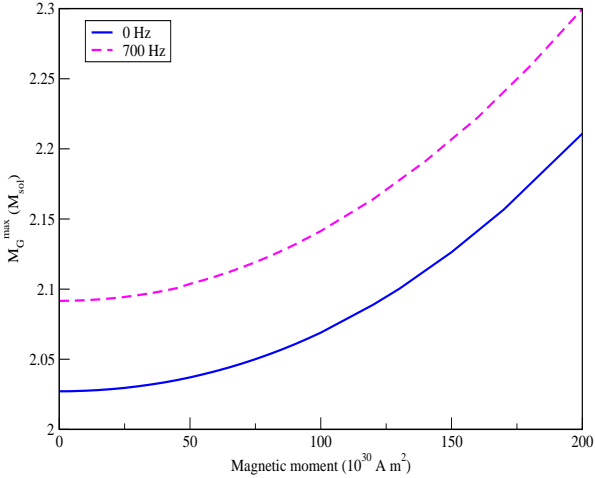}
        \caption{Maximum gravitational mass as function of magnetic
          moment for static (0~Hz) and rotating (700~Hz)
          configurations, with inclusion of magnetisation and magnetic
          field dependence in the EoS.}
    \label{fig:mmmgmax_rot}
  \end{center}
\end{figure}

Finally, we computed rotating configurations along a sequence of
constant magnetic dipole moments. For the moment the observed
magnetars all rotate very slowly with periods of the order of seconds,
see \cite{Mereghetti}, mainly because the strong magnetic fields
induce a very rapid spin-down. This means that the fast rotating
configurations do not have any realistic observed counterpart for the
moment, and we perform this investigation mainly for curiosity.  As
obtained in the static case, the maximum gravitational mass was found
to increase with the magnetic dipole moment $\cal{M}$. In particular,
we chose a sequence of neutrons stars rotating at 700~Hz, close to the
frequency of the fastest known rotating pulsar, which rotates at 716
Hz (\cite{Hessels}). In Fig.~(\ref{fig:mmmgmax_rot}) we see the same
behaviour for both cases: the maximal mass increases with the magnetic
field and, although it is not shown in the figure, the effects of
magnetisation or inclusion of the magnetic field are very small, as in
the non-rotating case.

\section{Conclusions}\label{s:conc}

In this work, we developed a self-consistent approach to determine the
structure of neutron stars in strong magnetic fields, relevant for the
study of magnetars. Starting from the microscopic Lagrangian for
fermions in a magnetic field, we derived a general expression for the
energy-momentum tensor of one fluid in presence of a non-vanishing
electromagnetic field.  Due to the perfect conductor assumption, the
electric field vanishes in the fluid rest frame, and therefore only
magnetisation and the magnetic field dependence of the equation of
state enter the final results. Equations for the star's equilibrium
are obtained as usual from the conservation of the energy-momentum
tensor coupled to Maxwell and Einstein equations. This consistent
derivation shows in particular that, as claimed by \cite{Blandford},
the equilibrium only depends on the thermodynamic equation of state
and magnetisation explicitly only enters Maxwell and Einstein
equations. This should answer some discussion in the recent literature
on the role of magnetisation, see e.g. \cite{Dexheimer} and
\cite{Potekhin}.

We have extended an existing axisymmetric numerical code to solve
these coupled equations and to obtain static and rotating neutron star
configurations in general relativity. Taking as an example the
equation of state of quark matter in the MCFL phase, we then
investigated the effect of inclusion of the dependence of the EoS on
the magnetic field as well as the magnetisation on the structure of
the neutron star. In contrast to the results obtained previously by
other authors, see~\cite{Paulucci, Dexheimer, Sinha}, we found that
the effect of inclusion of the magnetic field dependence on the EoS
does not change significantly the star's structure. In particular, we
have shown that the maximum mass of the neutron star is only slightly
modified even for the strongest magnetic fields considered, well above
those values which we could consider as realistic from present
magnetar observations. The difference to previous results arises due
to the fact that in these works the isotropic TOV equations were used
to solve for the star's structure, whereas the magnetic field causes
the star to deviate from spherical symmetry considerably. We hereby
confirm, within our fully consistent model, the conjecture by
\cite{Broderick}, who argued that the influence of the magnetic field on the EoS starts to become important only for values where the star has already taken a
toroidal shape, comparing the model by \cite{Cardall} with
calculations of the magnetic field dependence of several realistic
EoS.

An obvious question is of course to which extent we could expect
higher field values with stronger effects induced via the magnetic
field dependence of the equation of state. It is the surface poloidal
magnetic field which is estimated from the observation of spin-down of
pulsars via electromagnetic radiation. In this work, we have studied
magnetised neutron star models with a purely poloidal magnetic
configuration, taking maximum field values well above those
observed. It has, however, been argued that differential rotation in
neutron stars could amplify the seed poloidal field resulting in the
generation of a toroidal field with higher absolute value than the
poloidal one. The effect of purely toroidal fields on the structure of
neutron stars has already been elaborately studied in~\cite{Kiuchi,
  Yasutake, Frieben, Yoshida}. The virial theorem allows to estimate a
maximum limit to the magnetic field in the interior, irrespective of
the configuration, from the observed poloidal surface fields,
indicating that the maximum values considered within this work are
probably not exceeded in realistic situations. Thus, although the
study of magnetised neutron star models with purely poloidal magnetic
field is not the most general one, it gives us a fairly good idea
about the effect of the maximum field on the stellar structure. Our
formalism can of course be extended to any magnetic field
configuration and in a future work this statement should be tested
quantitatively.

In addition, as the main aim of the present work was to build
consistent models and to show the numerical applicability of the
formalism, we have tested it and computed neutron star models only
with one particular equation of state. In principle, it could be that
another model shows stronger effects. For hadronic models, as shown in
\cite{Broderick2000}, the effect of Landau quantization on the proton
Fermi energy starts to become considerable for field values of the
order $5 \times 10^{18}$ G, thus too high to influence the magnetar
structure. Another point is that we have not taken into account the
contribution of anomalous magnetic moments. A simple estimate of
energetics shows, however, that the magnetic interaction energy for
(vacuum) nucleonic magnetic moments is of the order $0.006 B_{10^{15}
  G}$ MeV, thus of the order of several MeV for fields of $10^{18}$
G. This is still very small compared with the energy scale of a high
density EoS, we therefore do not expect a very strong effect on the
global structure, either. To be more quantitative, in a typical
hadronic RMF model, with Landau quantization and anomalous magnetic
moments, the magnetisation is about ten times larger than in our quark
matter example, see~\cite{Rabhi2014}, and the effects should thus be
only slightly larger.

The possibility of a ferromagnetic instability in nuclear matter has
been discussed, too, see e.g. \cite{Vidaurre, Maruyama}. The general
opinion was that this instability is unphysical since no such
behaviour is observed for nuclear matter in heavy ion collisions nor
in microscopic nuclear matter calculations starting from the
fundamental nucleon-nucleon interaction. Ferromagnetic behaviour in
neutron star matter --quark or hadronic-- is, however, not
unambiguously refuted, see for instance the recent works of
\cite{Diener} and \cite{Tsue}. In the context of strong magnetic
fields inside a neutron star, ferromagnetism could considerably
enhance the effect of the magnetic field on the equation of state and
the corresponding neutron star structure. The magnetic catalysis
effect, modifying hadronic masses in a strong magnetic field,
see~\cite{Haber}, could additionally influence the equation of state,
too.

To close, let us mention that our code is part of \cite{Lorene} and
thus publicly available. It can therefore serve as a basis for testing
consistently the magnetic field effect within different types of
models on neutron star structure in the future in order to give a more
complete answer whether the magnetic field dependent modification of
the equation of state has to be considered quantitatively upon
describing magnetars.

\section*{Acknowledgements}
We gratefully acknowledge valuable discussions with Veronica
Dexheimer, \'Eric Gourgoulhon and Constan\c{c}a
Provid\^encia. Constan\c{c}a Provid\^encia also gave important remarks
upon carefully reading the manuscript. This work has been partially
funded by the SN2NS project ANR-10-BLAN-0503, the ``Gravitation et
physique fondamentale'' action of the Observatoire de Paris, and the
COST action MP1304 ``NewComsptar''.

\appendix

\section{Derivation of thermally averaged energy-momentum
  tensor} \label{a:App1} 

The following method follows closely that employed in
\cite{Ferrer2010}. For the further developments, we will distinguish
between the fermionic contribution to $T^{\mu\nu}$,
\begin{eqnarray}
  T^{\mu\nu}_f &=& \frac{1}{2}\bar{\psi}(\gamma^\mu \partial^\nu +
  \gamma^\nu  \partial^\mu) \psi - \frac{1}{2} ( j^\mu A^\nu + j^\nu A^\mu)
  \nonumber \\ &&
  -
  g^{\mu\nu}\bar{\psi}(x)(D_\mu \gamma^\mu + m)\psi(x)~,\nonumber
\end{eqnarray}
and the purely electromagnetic one, 
\begin{displaymath}
  T^{\mu\nu}_{EM} = -\frac{1}{\mu_0} (F^{\mu\alpha} F_{\alpha}^{\,\nu} +
  g^{\mu\nu} \frac{1}{4} F_{\alpha\beta} F^{\alpha\beta})~.
\end{displaymath} 
Within the former expression we have introduced the electromagnetic current, 
\begin{displaymath}
j^\mu = \frac{\partial {\mathcal L}}{\partial A_\mu} = i q \,\bar\psi \gamma^\mu
\psi~.
\end{displaymath}

The electromagnetic part is treated as completely classical here and
we will assume in addition that the electromagnetic fields are
constant over the averaging volume.  It is then trivial to show that
for the purely electromagnetic part we obtain simply
\begin{equation}
  \langle  T_{EM}^{\mu\nu} \rangle = -\frac{1}{\mu_0} (F^{\mu\alpha}
  F_{\alpha}^{\,\nu} + g^{\mu\nu} \frac{1}{4} F_{\alpha\beta} F^{\alpha\beta}) ~.
\end{equation}

The fermionic part is a little bit more complicated to evaluate.  Let
us first compute some derivatives of the partition function related to
thermodynamic quantities. We follow here the standard formalism, as
employed e.g.~\cite{Ferrer2010}. We will thereby factorise the
partition function as $Z = Z_{EM} Z_f$ which is possible since we
treat the electromagnetic part as classical and constant. It can thus
be taken out of the integral. The first derivative will be $\partial
Z_f/\partial \beta$. To that end we perform the variable
transformation $\lambda \to \lambda' \beta$. The partition function
then reads
\begin{eqnarray}
  Z_f =  \, \int {\mathcal D} \psi {\mathcal D} \bar\psi
  \exp\left(- \beta \int_0^1 d\lambda' \int d^3 x\times \right.\nonumber\\
  \left. \bar{\psi} \left( (\frac{i}{\beta} \partial_{\lambda'} - i q  A_0 + i \mu
      \hat{Q} ) \gamma^0 + D_i \gamma^i+ m \right)\psi\right) ~, 
\end{eqnarray}
and we obtain upon inverting the variable transform
\begin{equation}
  \frac{\partial Z_f}{\partial \beta} = \int {\mathcal
    D} \psi {\mathcal D} \bar\psi \exp(\tilde S_f) \frac{1}{\beta}\int_0^\beta
  d\lambda \int d^3 x ({\mathcal L}_f  - \mu \hat{n} - \bar{\psi} \partial_0
  \gamma^0 \psi)~.
  \label{eq:dzdb}
\end{equation}
Reintroducing the electromagnetic part of the partition function we can
rewrite the above expression as a thermal expectation value
\begin{eqnarray}
  \frac{1}{\beta V} \beta \frac{\partial \ln Z_f}{\partial \beta}& =&
  \frac{1}{\beta V} \frac{1}{Z} \int {\mathcal
    D} \psi {\mathcal D} \bar\psi \exp(\tilde S) \times \nonumber \\
  && \int_0^\beta
  d\lambda \int d^3 x ({\mathcal L}_f  - \mu \hat{n} + \bar{\psi} \partial_0
  \gamma^0 \psi) \nonumber \\ &=& -\langle {\mathcal L}_f - \mu \hat{n} +
  \bar{\psi} \partial_0 \gamma^0 \psi \rangle ~.
  \label{eq:domegadb}
\end{eqnarray}
In the above expression we recognise the grand canonical potential, 
\begin{equation}
  \Phi_f = -\frac{1}{\beta V} \ln Z_f~,
  \label{eq:omega}
\end{equation}
with the following well-known thermodynamic relations
\begin{subequations}
  \begin{eqnarray}
    n &=& -\frac{\partial \Phi_f}{\partial \mu}\\
    \frac{1}{\beta V} \beta \frac{\partial \ln Z_f}{\partial\beta} &=& - \Phi_f + T
    \frac{\partial \Phi_f}{\partial T} = -\Phi_f - T s \nonumber \\
    &=& -\varepsilon + \mu n\\
    \frac{1}{\beta V}  V \frac{\partial \ln Z_f}{\partial V} &=& -\frac{\partial
      \Phi_f \, V}{\partial V} = p ~.
  \end{eqnarray}
\end{subequations}
We have introduced here the number density $n$, the entropy density $s$,
pressure $p$ and the energy density $\varepsilon$.

Following the same procedure with the variable transform $x^i \to L_i
x^i$ and the volume given by $V = L_1 L_2 L_3$ we can show that
\begin{equation}
  \frac{1}{\beta V}  V \frac{\partial \ln Z_f}{\partial V} = \langle \bar\psi
  \gamma^i \partial^i \psi + {\mathcal L}_f\rangle~.
  \label{eq:zvolume}
\end{equation}

The next step will be the derivative with respect to the
electromagnetic field strength tensor.  Assuming a perfect conductor,
the electric field vanishes in the FRF and only the magnetic field
$b_\mu $ is nonzero.  In order to be precise we will, under the
assumption of constant electromagnetic fields over the averaging
volume, take
\begin{equation}
  A_\mu = - \frac{1}{2} F_{\mu\nu}(y) (x-y)^\nu
  \label{eq:amu}
\end{equation}
at some given point $y$, where $F_{\mu\nu}$ is given by
Eq.~(\ref{eq:deffmunu}). 

It can be shown that
\begin{equation}
  \frac{\partial \tilde S_f}{\partial b^\tau} = - \frac{1}{2}
  \epsilon_{\mu\nu\tau\sigma} u^\sigma 
  \int_0^\beta d\lambda \int d^3x (x-y)^\nu j^\mu~.
\end{equation}
Hence,
\begin{eqnarray}
  \frac{1}{\beta V} \frac{\partial \ln Z_f}{\partial
    b^\tau} &=& 
  \frac{1}{\beta V} \frac{1}{Z_f} \int {\mathcal
    D} \bar\psi {\mathcal D} \psi \exp(\tilde S_f) \times\nonumber \\ &&
  \int_0^\beta d\lambda \int d^3 x (
  - \frac{1}{2}
  \epsilon_{\mu\nu\tau\sigma} u^\sigma (x-y)^\nu j^\mu)   \nonumber\\
  &=&  - \frac{1}{2}
  \epsilon_{\mu\nu\tau\sigma} u^\sigma \langle (x-y)^\nu j^\mu \rangle~.
  \label{eq:mtau}
\end{eqnarray}
The above equation gives the magnetisation vector,
\begin{subequations}
  \begin{eqnarray}
    m_\tau &=& -\frac{\partial \Phi_f}{\partial b^\tau} = \frac{m}{b} b_\tau
    \label{eq:mtaudef} \\
    & =& -\frac{1}{2} \epsilon_{\mu\nu\tau\beta} u^{\beta} \langle
    (x-y)^\nu j^\mu\rangle~,
  \end{eqnarray}
\end{subequations}
As can be seen here, in isotropic matter, the magnetisation is aligned
with the magnetic field, and the polarisation components vanish due to
the perfect conductor assumption in the FRF.  From
Eq.~(\ref{eq:mtaudef}) the magnetisation tensor, $M_{\mu\nu}$, can be
calculated via Eq.~(\ref{eq:defmmunu}). It is obviously antisymmetric,
$M_{\mu\nu} = -M_{\nu\mu}$, as it should be and fulfils the
requirement that its divergence gives the expectation value of the
current, too.

Using Eq.~(\ref{eq:amu}), we can write in addition
\begin{equation}
  \langle A^\mu j^\nu \rangle = - \frac{1}{2} \epsilon^{\mu\rho\alpha\beta}
  b_\alpha u_\beta \langle (x-y)_\rho j^\nu \rangle~.
\end{equation}
Inserting Eq.~(\ref{eq:mtaudef}) into Eq.~(\ref{eq:defmmunu}) and
using the antisymmetry of $F_{\mu\nu} = - F_{\nu\mu}$ we obtain
finally
\begin{equation}
  -\frac{1}{2} \langle A_\mu j_\nu + A_\nu j_\mu \rangle = \frac{1}{2}
  (F_{\mu}^{\;\rho} M_{\rho\nu} + F_{\nu}^{\; \rho} M_{\rho\mu})~. 
\end{equation}

Let us now come back to the energy-momentum tensor.  Neglecting for
the moment the coupling to the electromagnetic field, we can show
that, in the FRF, the non-diagonal elements of the energy-momentum
tensor in homogeneous isotropic matter have a vanishing expectation
value. This can be seen by writing $\psi(x)$ in a Fourier basis with
respect to momentum $p^\mu$. The derivative $i \partial_\mu$ adds a
factor $p^\mu$ and the non-diagonal elements become the product of an
even and an odd function in $p^\mu$. This product vanishes upon
integration over $d^4 p$. Formulated in another way we use here the
fact that the movement of one homogeneous isotropic fluid is described
by one vector, its four-velocity $u^\mu$. As a consequence the
energy-momentum tensor necessarily has the following structure
\begin{equation}
  \langle T^{\mu\nu}_f \rangle = a_1 \,u^\mu u^\nu + a_2 \,g^{\mu\nu} ~,
\end{equation} 
with two scalar functions $a_{1,2}$ which can be evaluated in the FRF as
\begin{eqnarray}
  a_2 &=& -(u\cdot u) \langle T^{ii}_f\rangle = -(u\cdot u) \frac{\partial \ln
    Z}{\beta \partial V} = -(u\cdot u) p \\
  a_1 &=& \langle T^{00}_f\rangle + \langle T^{ii}_f\rangle  =
  \varepsilon + p~.
\end{eqnarray}
$p$ and $\varepsilon$ are the pressure and energy density of
the system, respectively. 

In the presence of an electromagnetic field, the eigenstates of
charged fermions are no longer momentum eigenstates. They become
quantized in the plane perpendicular to the direction of the magnetic
field $b_{\mu}$, see~\cite{Landau}.  It can again be shown, e.g. with
the help of an explicit representation of the spinor fields in
presence of a magnetic field, that in the FRF the thermal average of
the non-diagonal elements of $T^{\mu\nu}_f$ vanish except those
involving explicitly the magnetic field. We thus obtain formally the
same result for the fermionic part of the energy-momentum tensor,
\begin{eqnarray}
  \langle T^{\mu\nu}_f \rangle&=& (\varepsilon + p) \; u^\mu u^\nu  + p\;
  g^{\mu\nu} \nonumber \\
  && + \frac{1}{2} (F^\nu_{\;\tau} M^{\tau\mu} +
  F^{\mu}_{\;\tau} M^{\tau\nu})~. 
\end{eqnarray}
Pressure and energy density are defined as derivatives of the
fermionic partition function in the FRF as before. Putting everything
together we obtain the final result, see Eq.~(\ref{eq:tmunu}).


\label{lastpage}
\end{document}